# Analysis of a Play by Means of CHAPLIN, the Characters and Places Interaction Network Software

A.C. Sparavigna[1], R. Marazzato[2]

[1]Department of Applied Science and Technology, Politecnico di Torino, Italy
[2]Department of Control and Computer Engineering, Politecnico di Torino, Italy

**Abstract:** Recently, we have developed a software able of gathering information on social networks from written texts. This software, the CHAracters and PLaces Interaction Network (CHAPLIN) tool, is implemented in Visual Basic. By means of it, characters and places of a literary work can be extracted from a list of raw words. The software interface helps users to select their names out of this list. Setting some parameters, CHAPLIN creates a network where nodes represent characters/places and edges give their interactions. Nodes and edges are labelled by performances. In this paper, we propose to use CHAPLIN for the analysis a William Shakespeare's play, the famous "Tragedy of Hamlet, Prince of Denmark". Performances of characters in the play as a whole and in each act of it are given by graphs.

**Keywords**: Literary experiments, Networks, Graph Visualization Software, Text Data Analysis

## 1. Introduction

As we have discussed in some recent papers [1,2], the analysis of social networks, which is largely used in several theoretical and applied sciences, is currently becoming important also in the study of literary works. In fact, the network analysis can provide quantitative results concerning persons, places and interactions, extracting them from the layout of plays and novels [3].

Several studies and projects for automatic analysis of texts exist. In Reference 3, for instance, the authors are proposing an algorithm to infer the social network from an input text, based on some personal and geographical names previously defined. Social relationships are stated by dialogues and locations at which characters are active. Another approach was proposed in Reference 4, based on the detection of face-to-face conversations. By means of their algorithms, the researchers took a systematic and wide look at a large corpus of texts [4]. This is the research also pursued by the Stanford "Literary Laboratory" too. One of the projects of this Laboratory is the study of plots in terms of network theory [5].

Any automatic text analysis, like the one we have proposed in [6], needs to determine what we are expecting to find in a novel or a play, that is, the main characters about whom the work is pivoting, their interactions and also places and motivations of their behaviors. In [6], we have discussed a software we implemented in Visual Basic: it is CHAPLIN, the CHAracters and PLaces Interaction Network tool, able of performing basic descriptive statistics, evaluated from correlation concepts. By means of it, the characters of a literary work can be extracted from a list of raw words. Using a friendly interface, the user selects their names out of this list. Setting some parameters, CHAPLIN creates a network where nodes are characters and places, and edges represent their interactions. Nodes and edges are labelled by performances. The whole process is automated, except the phase in which human intervention is needed in recognizing characters and places. The result of the processing is a readable graph of a character/place network.

Here, we are proposing an example of using CHAPLIN, by applying it to the analysis a William Shakespeare's play, the famous "Tragedy of Hamlet, Prince of Denmark". After a short discussion of CHAPLIN, the graphs for the overall play and for each act of it will be displayed.

## 2. The CHAPLIN software

This software was developed to help automating the process of extracting, from a literary text, a meaningful graph representing the network of characters and places displayed in it. As explained in [6], this tool is not completely automated because a certain amount of human intervention is asked for recognizing basic characters and places. The required input is a set of text files stored in a folder, containing the acts/chapters of the literary work to process. This content can be viewed as a set:

$$T = \{C_i ; i \in (1,\ldots,c)\} \quad (1)$$

First, the tool automatically extracts a subset of the complete set of words appearing in T, by means of a suitable function X, which at the same time splits each text into single words and selects them under some reasonable constraint B, such as "take only words long more than...", "take only capitalized words", compositions of the previous and so on; let's name this set as "raw words":

$$W = \{w_j ; j \in (1,\ldots,w)\} = X(T, B) \quad (2)$$

In this manner, W contains the names of all characters and places of the work being processed, mixed with many other words. The user is then asked recognizing the names of characters and places, so generating the set

$$N = \{n_k ; k \in (1,\ldots,n)\} \quad (3)$$

As discussed in [6], CHAPLIN considers that to each name it could correspond a set of variants. In fact, a certain name can vary in its form or even appears to be quite different from the main form, for linguistic and/or narrative reasons, such as case and number inflection, epithets and so on. Therefore CHAPLIN involves finding the occurrences of each variant, grouping them together under the corresponding name. Then, a frequency distribution of names and an interaction matrix are produced:

$$\begin{aligned} F &= F(n_k) \\ I &= I(n_k, n_h) \end{aligned} \quad (4)$$

Each element of I represents how tight the character $n_k$ is linked to $n_h$; the numerical value is based on the values given by a proximity function:

$$\Pi(\Delta) : N \rightarrow [0,1] \subset R \quad (5)$$

for each couple of occurrences of $(n_k, n_h)$. In (4), the values of F and I are normalized to their maximum value, so that

$$\begin{aligned} F(n_k) &\in [0,1] \subset R \,\forall (n_k) \\ I(n_k, n_h) &\in [0,1] \subset R \,\forall (n_k, n_h) \end{aligned} \quad (6)$$

In a written work, it could appear a lot of secondary characters and places; in addition to that, some of them could be very weakly linked to each other. In order to avoid complex or nearly meaningless graphs, it is possible to select only the most relevant ones, by setting threshold values for the elements of both I and F. A graph is then created, where each node is carrying a name and its narrative strength F. Edges are linking the names, with their relative narrative interaction I. Let us note that orphan nodes, representing meaningful characters or places with scarce connection to others, can occur in the graph.

CHAPLIN is implemented in VB.NET. The tool has an interface for users which is including basic commands. It contains commands which are used to extract raw words and the related lists aiding the user in selecting raw words into names. As discussed in [6], the software gives a DOT script graph in a GV file [7], which can be rendered by means of the free open source software Graphviz [8].

## 3. An example: Shakespeare's Hamlet

Let us show here an example of using CHAPLIN on a play. This is the Shakespeare's play "The Tragedy of Hamlet, Prince of Denmark". The threshold parameters for the processing of nodes and edges are 0.15 for the character strength and 0.15 for the link strength. In this play we have not considered the places as relevant, and therefore we have not used them. The Figure 1 shows the result from processing the play as a whole. The Figures 2-6 are referring to each act of the play.

The protagonist is Hamlet, son of the recently deceased king of Denmark (the Ghost appearing in the play), and nephew of King Claudius, his father's brother and successor. Claudius hastily married King Hamlet's widow, Gertrude, Hamlet's mother. The other female character is Ophelia, who loves Hamlet.

From the Figure 1 we can notice the following:
 Many secondary characters do not appear in the graph.
 The protagonist of the play is Hamlet, and he obtains the best score as character; the second score is obtained by Claudius. In Ref.2, we had already seen that Claudius is the other "hub" of the play's network of characters. And in fact, Hamlet and Claudius are the nodes having the larger numbers of edges incident to them.
 The starkest link connects Hamlet to his friend Horatio. This is due to the fact that Horatio is present through most of the major scenes of the play.

In the Figures 2-6, we can see the graphs obtained by analyzing each single act. In this manner we have quantified the specific role of each character in each act.

## 4. Discussion

The CHAracters and PLaces Interaction Network (CHAPLIN) software gives the basic descriptive statistics and correlations, which are present in the play. Using it, we have the frequency of characters and the relative performances of their interactions. The whole process is automated, except the phase in which human intervention is needed for choosing names. The results are well readable graphs showing networks where characters are plotted as nodes, and the narrative connections between them correspond to edges. The importance of each character and the strength of links are represented by means of a score varying between 0 and 1.

Let us note that, besides characters and places, any other word in the text can be chosen from the list of raw words extracted by the tool. For instance, a word which is a leitmotiv of the text can be considered as well, as we did in [9], in an analysis of Dante's Divine Comedy. Therefore, leitmotivs can be added as nodes of the network that CHAPLIN can produce from the literary text. This will be discussed in a future paper.

## 5. Acknowledgement

The name we gave to this software is clearly inherited from that of Sir Charles Spencer Chaplin. The first version of software was released on Feb 7, 2014, exactly 100 year after the first appearance of the Tramp (Charlot). It is then an homage we paid to one of the most relevant persons of the seventh art.

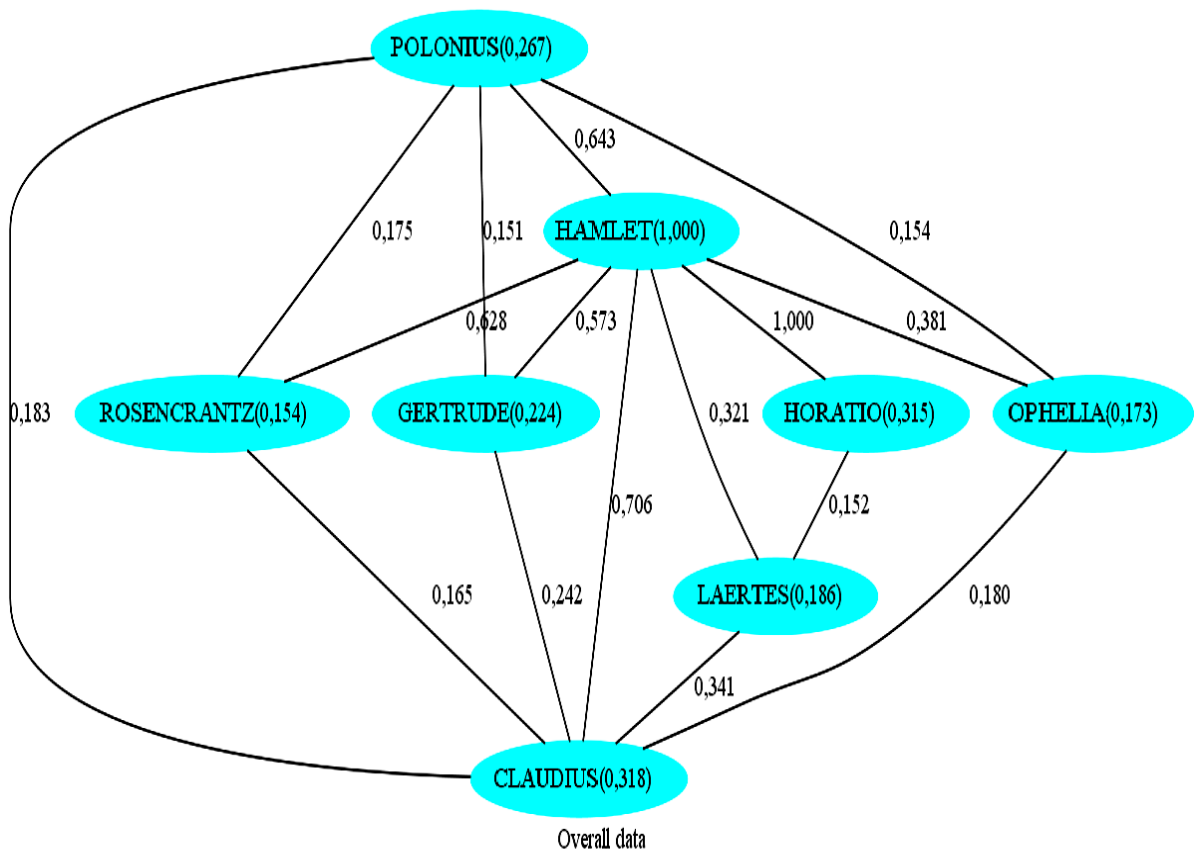

HAMLET: F=1,000
CLAUDIUS: F=0,318
HORATIO: F=0,315
POLONIUS: F=0,267
GERTRUDE: F=0,224
LAERTES: F=0,186
OPHELIA: F=0,173
ROSENCRANTZ: F=0,154

HAMLET—HORATIO: I=1,000
HAMLET—CLAUDIUS: T=0,706
POLONIUS—HAMLET: I=0,643
HAMLET—ROSENCRANTZ: I=0,628
HAMLET—GERTRUDE: I=0,573
HAMLET—OPHELIA: I=0,381
LAERTES—CLAUDIUS: I=0,341
HAMLET—LAERTES: I=0,321
GERTRUDE—CLAUDIUS: I=0,242
POLONIUS—CLAUDIUS: I=0,183
OPHELIA—CLAUDIUS: I=0,180
POLONIUS—ROSENCRANTZ: I=0,175
ROSENCRANTZ—CLAUDIUS: I=0,165
POLONIUS—OPHELIA: I=0,154
HORATIO—LAERTES: I=0,152
POLONIUS—GERTRUDE: I=0,151

**Figure 1:** This is the graph obtained by CHAPLIN, concerning the play as a whole. We can see that many secondary characters do not appear in it. The protagonist of the play is Hamlet, and he obtains the best score as character; the second score is obtained by Claudius. In Ref.2, we had already seen that Claudius is the other "hub" of the play's network of characters (Hamlet and Claudius are the nodes having the larger numbers of edges incident to them). The starkest link connects Hamlet to his friend, Horatio. This is due to the fact that Horatio is present through most of the major scenes of the play.

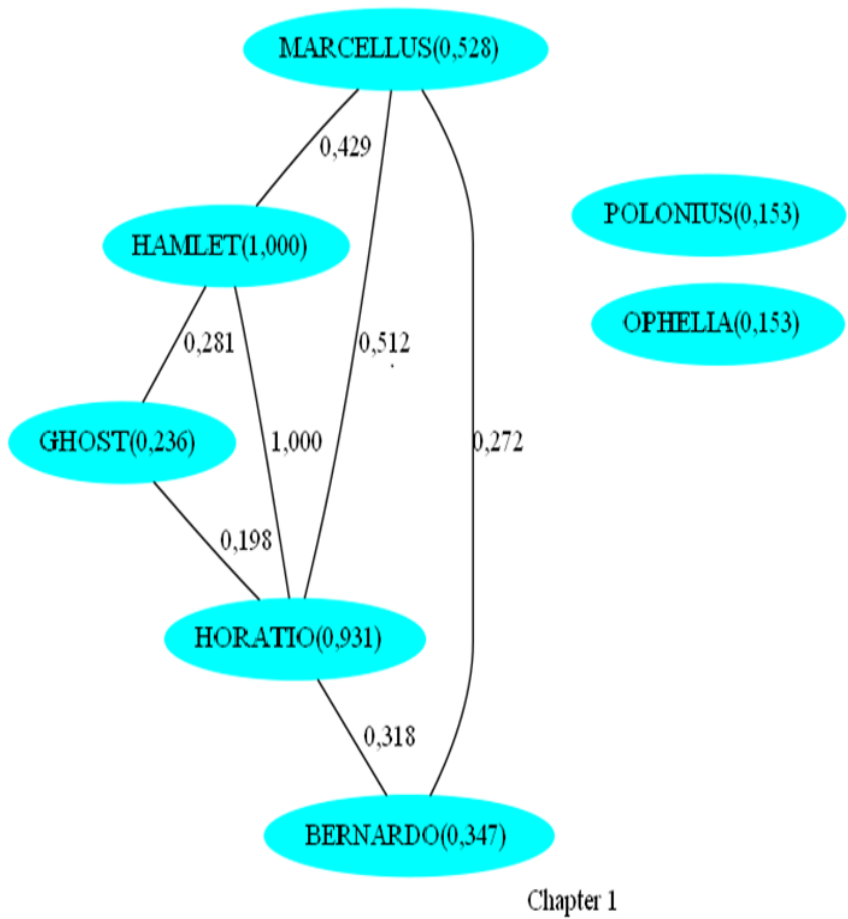

HAMLET: F=1,000
HORATIO: F=0,931
MARCELLUS: F=0,528
BERNARDO: F=0,347
GHOST: F=0,236
POLONIUS: F=0,153
OPHELIA: F=0,153

HAMLET—HORATIO: I=1,000
MARCELLUS—HAMLET: I=0,429
MARCELLUS—HORATIO: I=0,512
HORATIO—BERNARDO: I=0,318
HAMLET—GHOST: I=0,281
MARCELLUS—BERNARDO: I=0,272
GHOST—HORATIO: I=0,198

**Figure 2:** This is the graph obtained by CHAPLIN, concerning the first act of the play. We can see the presence of two "orphan" characters. This means that the strength of their related links is lower than the chosen threshold value of 0.15, whereas their character strength is above the threshold. Hamlet obtains the best score as character.

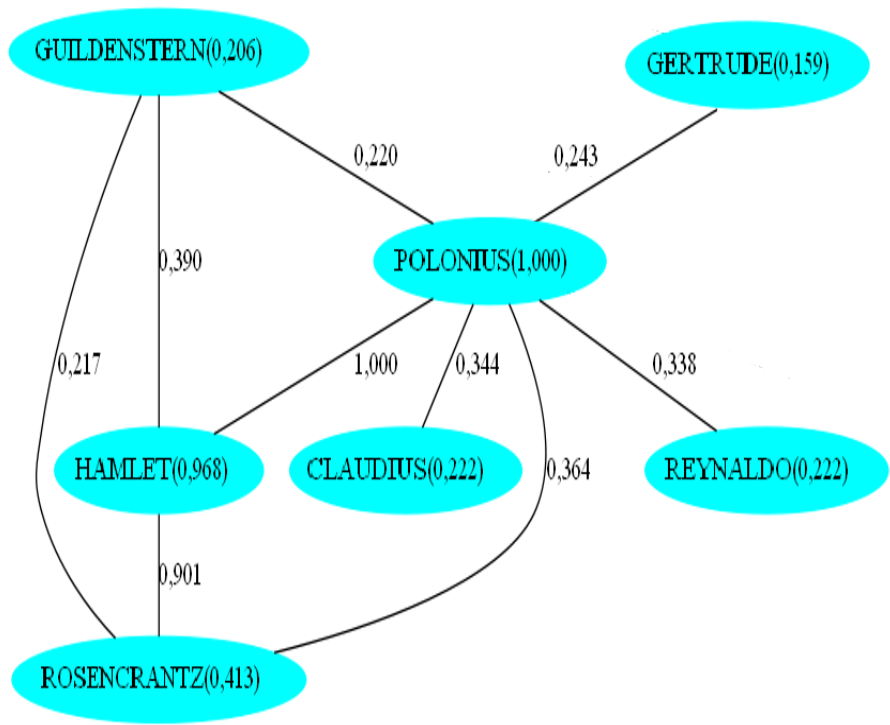

POLONIUS: F=1,000
HAMLET: F=0,968
ROSENCRANTZ: F=0,413
REYNALDO: F=0,222
CLAUDIUS: F=0,222
GUILDENSTERN: F=0,206
GERTRUDE: F=0,159

POLONIUS—HAMLEL: I=1,000
HAMLET—ROSENCRANTZ: I=0,901
GUILDENSTERN—HAMLET: I=0,390
POLONIUS—ROSENCRANTZ: I=0,364
POLONIUS—CLAUDIUS: I=0,344
POLONIUS—REYNALDO: I=0,338
POLONIUS—GERTRUDE: I=0,243
GUILDENSTERN—POLONIUS: I=0,220
GUILDENSTERN—ROSENCRANTZ: I=0,217

**Figure 3:** This is the graph obtained by CHAPLIN, concerning the second act of the play. Here Polonius obtains the best score as character. The best link is between Polonius and Hamlet. Polonius is chief counselor of the king, and the father of Laertes and Ophelia. This character connives with Claudius to spy on Hamlet.

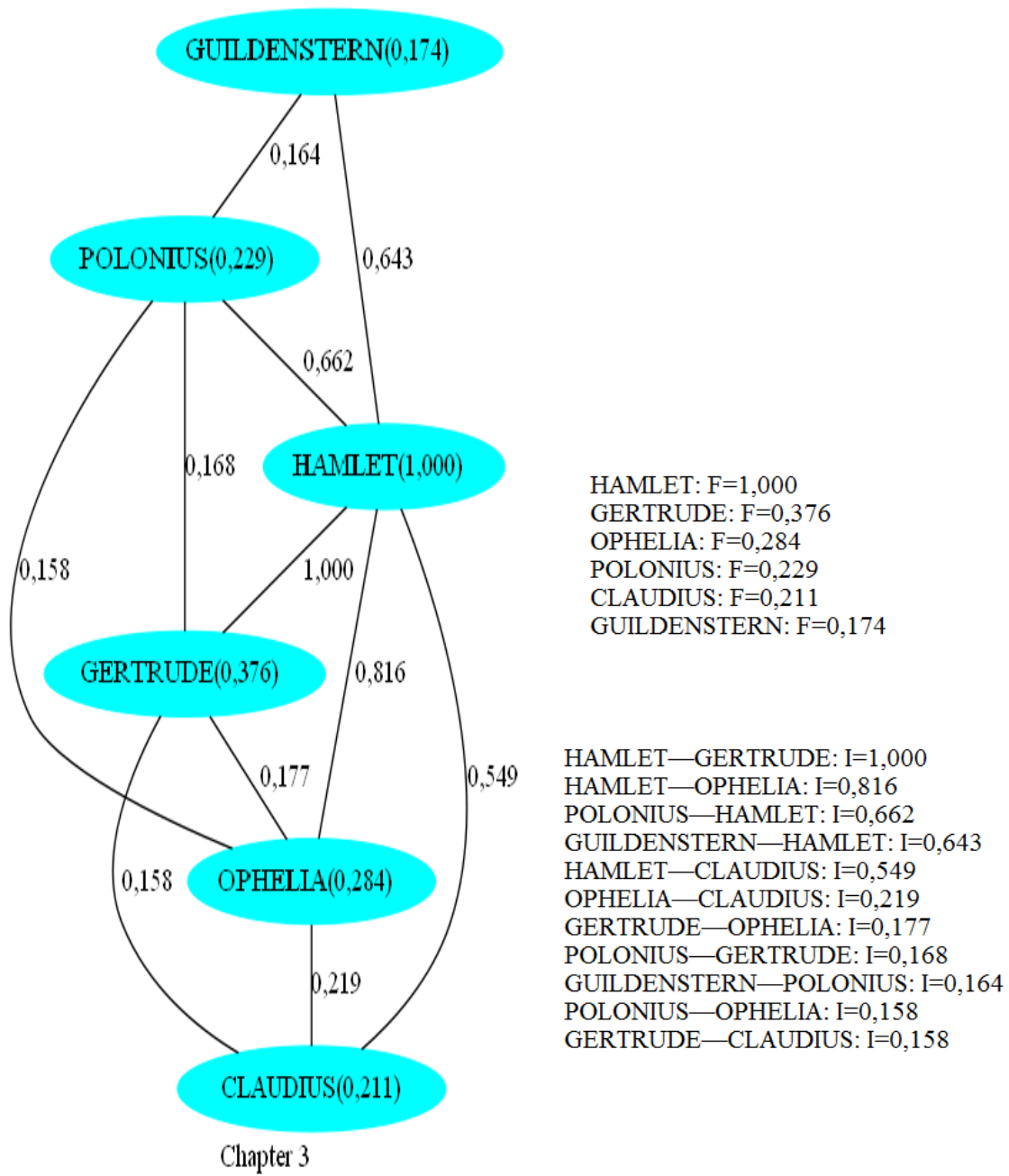

**Figure 4:** This is the graph obtained by CHAPLIN from the third act of the play. In this act, Hamlet obtains the best score as character. The second score is obtained by his mother Gertrude. The best link is between Hamlet and Gertrude. Also important in this act is Ophelia.

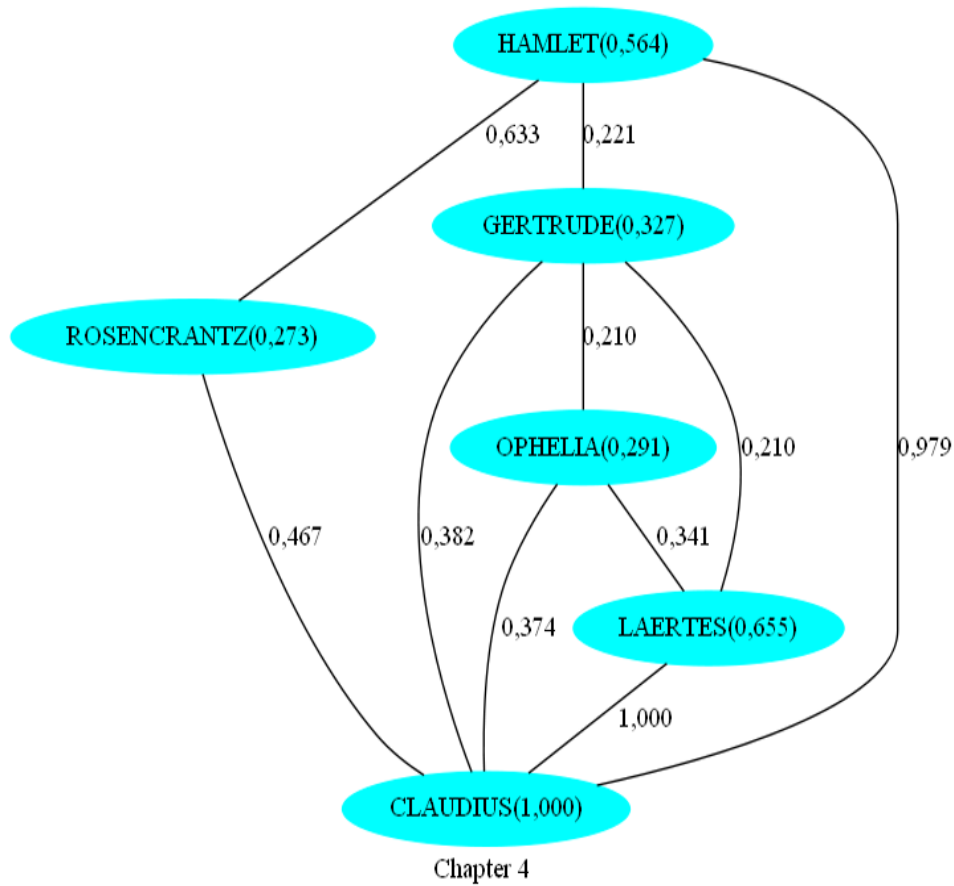

CLAUDIUS: F=1,000
LAERTES: F=0,655
HAMLET: F=0,564
GERTRUDE: F=0,327
OPHELIA: F=0,291
ROSENCRANTZ: F=0,273

LAERTES—CLAUDIUS: I=1,000
HAMLET—CLAUDIUS: I=0,979
HAMLET—ROSENCRANTZ: I=0,633
ROSENCRANTZ—CLAUDIUS: I=0,467
GERTRUDE—CLAUDIUS: I=0,382
OPHELIA—CLAUDIUS: I=0,374
OPHELIA—LAERTES: I=0,341
HAMLET—GERTRUDE: I=0,221
GERTRUDE—OPHELIA: I=0,210
GERTRUDE—LAERTES: I=0,210

**Figure 5:** Graph obtained by CHAPLIN, concerning the fourth act of the play. Claudius, the king, obtains the best score as character. Note that in this act, Claudius is the "hub" of the network, having the largest number of edges incident to his node.

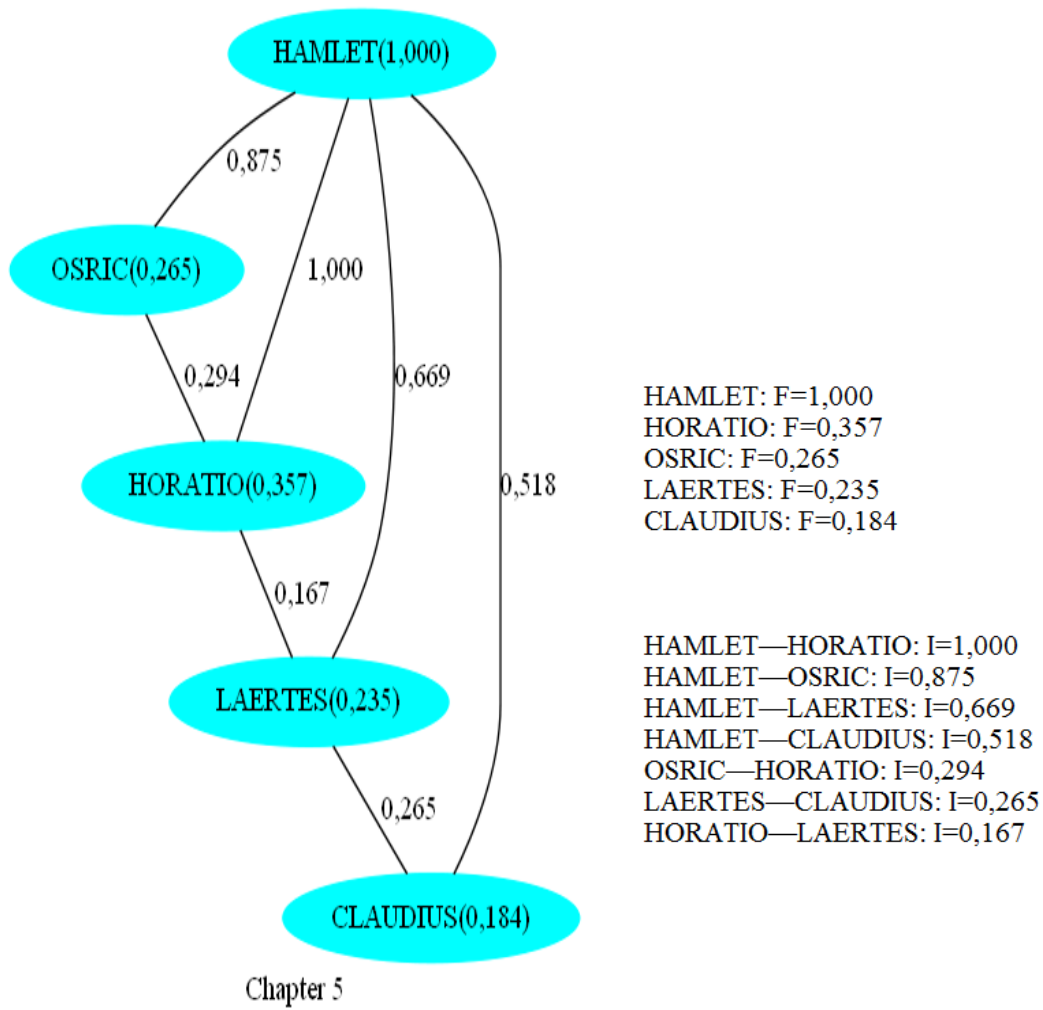

**Figure 6:** Graph obtained by CHAPLIN from the fifth act of the play. Hamlet is now obtaining the best score. This act opens with Hamlet and Horatio speaking with a gravedigger. Hamlet realizes that man's accomplishments are transitory, the skull of Yorick evoking his famous monologue on mortality.